\documentclass[lettersize,journal]{IEEEtran}
\usepackage{amsmath,amsfonts}
\usepackage{algorithmic}
\usepackage{algorithm}
\usepackage{array}
\usepackage[caption=false,font=normalsize,labelfont=sf,textfont=sf]{subfig}
\usepackage{textcomp}
\usepackage{stfloats}
\usepackage{url}
\usepackage{verbatim}
\usepackage{graphicx}
\usepackage{cite}
\usepackage{makecell}
\usepackage{subfig}
\begin{document}

\title{IEEE 802.11be Wi-Fi 7: Feature Summary \\ and Performance Evaluation}

\author{Xiaoqian Liu, Yuhan Dong, Yiqing Li, Yousi Lin and Ming Gan

\thanks{Xiaoqain Liu, Yousi Lin, and Ming Gan are with Huawei Technologies Co., Ltd, Shenzhen, China.}
\thanks{Xiaoqain Liu and Yuhan Dong are with Shenzhen International Graduate School, Tsinghua University, Shenzhen, China.}
\thanks{Yiqing Li is with Research Institute of China Telecom Corporation Limited.}

}



\maketitle

\begin{abstract}
As emerging applications demand increasingly higher throughput, IEEE standard 802.11be -- Extremely High Throughput (EHT), also known as Wi-Fi 7, was published on July 22, 2025. It can be used to meet the demand for the throughput of 4K/8K videos up to tens of Gbps and low-latency video applications such as virtual reality (VR) and augmented reality (AR).
Wi-Fi 7 not only scales Wi-Fi 6 with doubled bandwidth, but also supports real-time applications, which brings revolutionary changes to Wi-Fi.
In this article, we start by introducing the main objectives and timeline of Wi-Fi 7 and then list the latest key techniques which promote the performance improvement of Wi-Fi 7.
Finally, we validate the most critical objectives of Wi-Fi 7 -- the potential up to 30 Gbps throughput and lower latency.
System-level simulation results suggest that by combining the new techniques, Wi-Fi 7 achieves 30 Gbps throughput and lower latency than Wi-Fi 6.


\end{abstract}


\section{Introduction}

\IEEEPARstart{I}{n} a globalized society, wireless connectivity is considered necessary. Wi-Fi technology is one of the greatest successes of this information technology age, and it has created well-known economic benefits for the global community.
Wi-Fi connects people around the world and helps create a wide variety of new industries and careers.
Since Wi-Fi technology was adopted in 1990s, its market share has been growing steadily for many years and has played a significant role in providing wireless data services \cite{b1}.
Billions of people, including homes and enterprises, increasingly rely on Wi-Fi as their primary access network, which carries the majority of data traffic in an ever-expanding variety of applications around the world.

As the demand for wireless services continues to grow, Wi-Fi is also constantly evolving in terms of functionality.
Video traffic is a major component of Wi-Fi traffic, With the emergence of 4k and 8k video technologies with uncompressed data rates up to 20 Gbps, Wi-Fi throughput requirements will continue to increase \cite{b2}.
In addition, many new applications today require not only ultra-high throughput, but also extremely low latency.
Among them, virtual reality (VR), augmented reality (AR), gaming, telecommuting and screen projection have strict requirements on latency, which is generally less than 5 ms \cite{b3}. 
In the digital industry, in order to make wireless communication replace wired communication, reliability is also a major concern. The accuracy of data packets transmission needs to be higher than 99.99\%.
To address these requirements, consumers will demand further Wi-Fi improvements.

In recent years, the IEEE 802.11 Working Group (WG) has continued to find solutions to improve performance \cite{b4}.
IEEE 802.11be Task Group (TGbe) has specified the scope of the new generation Wi-Fi, i.e., enabling new medium access control (MAC) and physical (PHY) layers modes to support a maximum throughput of at least 30 Gbps \cite{PAR}.
At the same time, the new protocol requires at least one mode of operation capable of improving the worst-case latency and jitter\footnote{It is important to note that 802.11be WG has not defined specific evaluation index for latency and jitter. Our proposal \cite{laten} presented a detailed analysis on this matter, which will be described in Section \ref{lat}.} \cite{PAR}.
These requirements have been driving the development of IEEE 802.11be extremely high throughput (EHT) in recent years to meet the peak throughput set by upcoming applications.
Since identifying the main challenges of EHT, key experts have been discussing at IEEE meetings, exploring some PHY and MAC technologies and defining the building blocks of the IEEE 802.11be standard to overcome these challenges.

In this article, we introduce the application scenarios of the next-generation Wi-Fi and the main objectives of IEEE 802.11be, and summarize the recently launched standardization timeline. 
We delved into the latest concrete features to be adopted in the 802.11be amendment and detailed the implementations and benefits of each feature.
In addition, we discuss two important and appealing issues -- latency and throughput. In particular, we provide system-level simulation results to show that the features defined by 802.11be can achieve the throughput of 30 Gbps and improve latency performance.
Altogether, this article aims to summarize Wi-Fi 7 and provide a guide for researchers and general audiences.

\section{Wi-Fi 7 Progress}
\begin{figure*}[!t]
	\centering
	\includegraphics[width=7.37in]{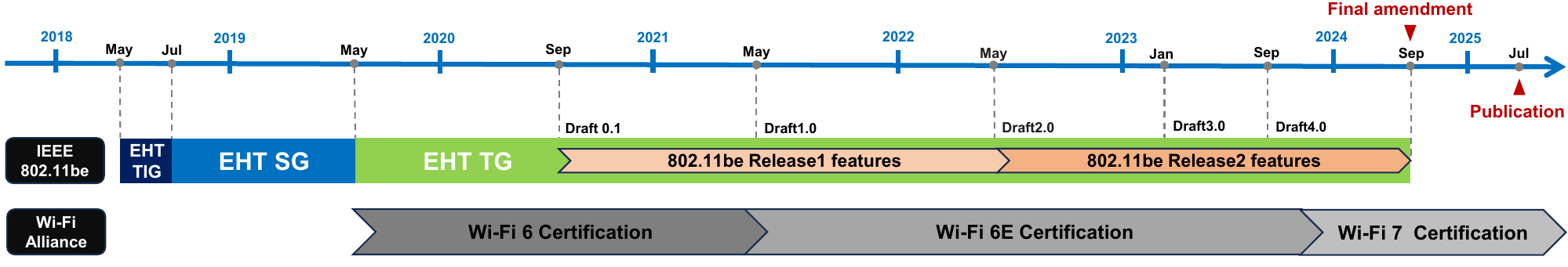}
	\caption{Illustration of recent Wi-Fi standardization and certification timeline.}
	\label{fig_1}
\end{figure*}

The standardization and certification timeline of Wi-Fi 7 is illustrated in Fig.~\ref{fig_1}.
The IEEE 802.11 WG approved the formation of a Topic Interest Group (TIG) in May 2018, and further established a Study Group (SG) and a Task Group (TG) in July 2018 and May 2019 respectively.
The primary objective of these groups is to ensure the rapid and stable progress of 802.11be EHT. 

In September 2020, 802.11be draft 0.1 was released, followed by draft 1.0 and draft 2.0 in May 2021 and May 2022, respectively. 
The IEEE 802.11be D7.0 draft, which is the final amendment, was approved by the IEEE 802 Executive Committee (802 EC) in September 2024, and the draft has now been frozen, marking the completion of the standard's development phase. Since early 2024, the Wi-Fi Alliance has also begun the certification process for Wi-Fi 7. In July 2025, IEEE standard 802.11be was officially published.
The 802.11be amendment spans five years and is expected to provide significant performance enhancements for Wi-Fi 7 users in all respects.

In the 802.11be development cycle, experts have focused their research interest on a set of features deemed high priority, depending on their market demand and complexity ratio. More information will be covered in the next section.

\begin{figure}[!t]
	\centering
	\includegraphics[width=3.5in]{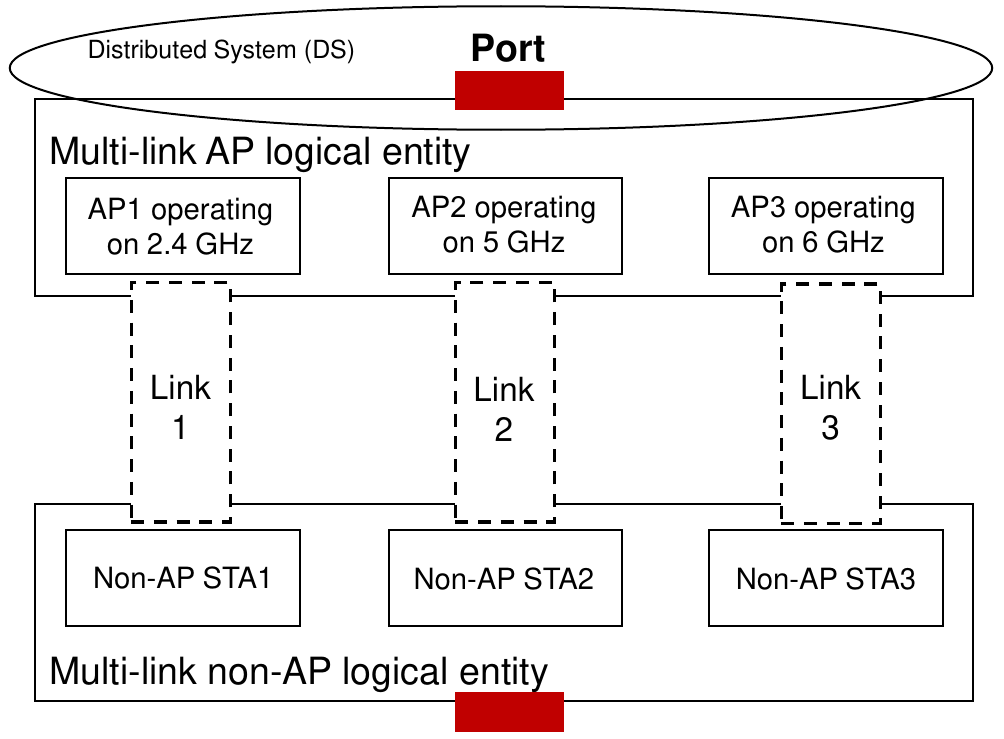}
	\caption{Illustration of multi-link architecture.}
	\label{fig_4}
\end{figure}

\section{Wi-Fi 7 Features}
In the 802.11be forum, experts from industry and academia have identified all the key features of Wi-Fi 7 which will be written into the consecutive drafts of the standard.
These features aim to support high-throughput and low-latency applications.
In the following, we detail the features that get the most attention.
\subsection{320 MHz Bandwidth}
Spectrum is the cornerstone of wireless communication, and the emergence of new generation of wireless communication technologies usually accompanies the use of new spectrum bands.
However, the obtainable bandwidth over 2.4 GHz and 5 GHz frequency bands cannot meet the throughput and latency requirements of emerging service such as AR/VR and online gaming \cite{b2}. 
EHT expands its bandwidth by aggregating spectrum across the 2.4 GHz, 5 GHz, and 6 GHz bands, achieving a total channel width of up to 320 MHz, which doubles the maximum nominal throughput compared to 802.11ax.

Note that Wi-Fi 7 supports large bandwidth, including 20/40/80/160/320 MHz, but only continuous case (i.e. single radio frequency chain).
When adopting large bandwidth, the simple replication of legacy preamble may worsen system performance.
To reduce it, the 320 MHz phase rotation with new tone plan and puncturing is applied on non-HT physical layer protocol data unit (PPDU) or non-EHT modulated preamble of EHT PPDU.


\subsection{Multiple Resource Unit (MRU)}
Orthogonal frequency division multiple access (OFDMA) is introduced to Wi-Fi in 802.11ax. It divides the channel resource into small time-frequency resource blocks, which are called resource units (RU), enabling multi-users to share channel resource and improving spectrum efficiency \cite{b5}.
RU is a relative concept, which means there are various sizes of RUs. In 802.11be, an RU can contain 26, 52, 106, 242, 484, 996, 2 $\times$ 996, 4 $\times$ 996 tones.
The maximum number of RUs varies according to the channel bandwidths (CBW). The more the number of RUs, the higher the multi-users processing efficiency and the higher the throughput.
The 20/40 MHz bandwidth tone plan is the same as the 802.11ax, while a new tone plan is designed for the 80/160/320 MHz bandwidth to make it more suitable for preamble puncture and easy to match spectrum mask \cite{b6}.
In 802.11ax, an access point (AP) can allocate only a single RU (SRU) to each station (STA). This restriction leads to network throughput decrease in the scenario with a small number of STAs.
In contrast, 802.11be supports the assignment of MRU per STA \cite{b6}. 
In addition, OFDMA and single user (SU) allow punctured transmission in 802.11be.


\subsection{4096-QAM}
To further enhance the peak rate, TGbe has adopted a higher-order modulation scheme. 
Compared with the 1024-QAM used in 802.11ax, the 4096-QAM is introduced in 802.11be, which increases the nominal rate by 20\% using the same coding rate \cite{b1}.
In 4096-QAM, a modulation symbol can carry 12 bits of information, which enables higher transmission efficiency.
But at the same time, the cost and complexity of devices that implement 4096-QAM are both high.
The demand for signal-to-noise ratio (SNR) at the receiver side is high, which is often achieved through strict error vector magnitude (EVM) requirement and beamformed transmission.

\subsection{1024 Aggregate Mac Protocol Data Units (A-MPDU)}
A-MPDU was introduced in the IEEE 802.11n Wi-Fi standard, which is a feature dedicated to enhancing Wi-Fi speed and efficiency \cite{bmpdu}.
In previous Wi-Fi standards (such as IEEE 802.11a/b/g), only one data frame could be transmitted at once. 
However, with the advancement of technology and the demand for higher transmission rates, there was a need to utilize wireless transmission capabilities more efficiently, leading to the introduction of frame aggregation technology.

In A-MPDU technology, the transmitter aggregates a series of MPDUs that have undergone 802.11 packet encapsulation, and separates them by MPDU delimiters.
The aggregated frame only retains one PHY header, which reduces the additional information of the PHY header for each MPDU transmission.
After receiving the A-MPDU, the receiver only needs to send a block acknowledgement (ACK), which will reduce the time consumed by channel contention, thereby improving system throughput.
Every generation of Wi-Fi standards after IEEE 802.11n has specified the number of MPDUs that can be aggregated by A-MPDU. In IEEE 802.11be standard, this number increased to 1024, which means a further enhancement in data throughput and transmission efficiency.




\subsection{Multi-Link Operations (MLO)}

The MLO is an important new feature of 802.11be \cite{MLOtcom}. 
Through MLO, APs and STAs will be provided with the capabilities to transmit and receive data from the same traffic flow over multiple radio interfaces.
Its adoption represents the development tendency towards multi-link communications.
TGbe has made changes in architecture, management functions, and power save to implement MLO and utilizes all available spectrum resources in 2.4 GHz, 5 GHz and 6 GHz.
Next, we explore the key features of MLO.


\subsubsection{\bf{Architecture}}
802.11be defines the multi-link device (MLD), which incorporates the traditional single-link AP or STA functionalities as internal affiliated entities across multiple links.
Both AP MLDs and STA MLDs are single devices with multiple physical radio interfaces.
However, it is noteworthy that there is only one MAC address per MLD, and individual parameters for each interface are not lost.
Fig.~\ref{fig_4} shows the MLD architecture, multiple connections can be set up between APs and STAs. 
MLD can be considered as a collaborative work scheme.
This scheme allows retransmission of packets on any link, which simplifies dynamic link switching and packet reassembly.


MLD is divided into two different transmission modes: simultaneous TX and RX (STR) mode and non-STR (NSTR) mode \cite{b8}.
STR MLD can transmit frames asynchronously over multiple links. This mode can be applied to both AP MLD and non-AP MLD and keep independent channel access in different links.
STR MLD can achieve higher throughput, but power consumption may increase due to simultaneous operation of multiple interfaces. Therefore, this mode must be followed by a power save mechanism, which will be described later.
NSTR MLD does not allow information to be transmitted over an idle interface while they are receiving through another and and can only be applied to non-AP MLDs.
This mode can avoid in-device coexistence (IDC) interference problems, but at the cost of reducing throughput.

\begin{table}[t]
	\renewcommand\arraystretch{1.4}
	\begin{center}
		\caption{Simulation setup parameters for PAR verification}
		\label{table_2}
		\setlength{\tabcolsep}{3 mm}{
			\begin{tabular}{|c|c|}
				\hline   \multicolumn{2}{|c|}{\textbf{Throughput verification}}  \\
				\hline   \rule{0pt}{8pt} \textbf{Simulation Setup Parameters} & \textbf{Default value}  \\   
				\hline   \rule{0pt}{8pt} Traffic type & Full buffer UL/DL traffic  \\ 
				\hline   \rule{0pt}{8pt} Packet size & 1500 Bytes per MPDU  \\ 
				\hline   \rule{0pt}{8pt} TX power & \makecell{\rule{0pt}{8pt} 20 dBm for AP MLD \\ 15 dBm for non-AP MLD}  \\ 
				\hline   \rule{0pt}{8pt} NSS & 8  \\   
				\hline   \rule{0pt}{8pt} CWmax & 15  \\ 
				\hline   \rule{0pt}{8pt} CWmin & 7  \\ 
				\hline   \rule{0pt}{8pt} AIFS & 34 $\mu$s  \\ 
				\hline   \rule{0pt}{8pt} The number of links & 2 (5 GHz \& 6 GHz)  \\ 
				\hline   \rule{0pt}{8pt} TXOP & 4.096 ms  \\ 
				\hline   \rule{0pt}{8pt} RTS/CTS & on  \\ 
				\hline   \multicolumn{2}{|c|}{\textbf{Low latency verification}}  \\
				\hline   \rule{0pt}{8pt} \textbf{Simulation Setup Parameters} & \textbf{Default value}  \\   
				\hline   \rule{0pt}{8pt} Traffic type & UL/DL traffic  \\ 
				\hline   \rule{0pt}{8pt} Traffic rate & 20-200 Mbps  \\ 
				\hline   \rule{0pt}{8pt} Packet size & 1500 Byte per MPDU	  \\   
				\hline   \rule{0pt}{8pt} MCS & 11  \\ 
				\hline   \rule{0pt}{8pt} NSS & 2 per STA  \\ 
				\hline   \rule{0pt}{8pt} Max aggregation & 256 MPDU per AMPDU  \\   
				\hline   \rule{0pt}{8pt} CWmax & 15  \\ 
				\hline   \rule{0pt}{8pt} CWmin & 7  \\ 
				\hline   \rule{0pt}{8pt} AIFS & 34 $\mu$s  \\ 
				\hline   \rule{0pt}{8pt} The number of links in MLO & 2  \\ 
				\hline   \rule{0pt}{8pt} Retry limit & 10  \\ 
				\hline   \rule{0pt}{8pt} MSDU life time & 20 ms  \\ 
				\hline   \rule{0pt}{8pt} RU size & \makecell{\rule{0pt}{8pt} 80 MHz: 242 tone \\ 160 MHz: 484 tone}  \\ 
				\hline   \rule{0pt}{8pt} AP MLD scheduling algorithm & Round Robin  \\ 
				\hline   \rule{0pt}{8pt} TXOP & 4.096 ms  \\ 
				\hline   \rule{0pt}{8pt} RTS/CTS & on  \\ 
				\hline
		\end{tabular} }
	\end{center}
\end{table}

\subsubsection{\bf{Discovery}}
In the discovery mechanism already defined in the 802.11 standard, the information about nearby APs is collected by the station passive/active scanning. 
Due to the introduction of MLD in 802.11be, the discovery mechanism needs to be updated.
To avoid excessive time consumption during the discovery process, 802.11be reuses reduced neighbor report (RNR) element, which can report information about different interfaces of the same AP MLD.
This method enables the station to directly detect the AP MLD, reducing the time from discovery to multi-link setup \cite{b7}.


\subsubsection{\bf{Power Save}}
Since the most major source of traffic in the information age is portable mobile devices, issues related to power consumption must be carefully considered.
The use of MLD results in the presence of at least two devices at the same time, which consumes twice more energy.
When the traffic is light, the device does not need to continuously listen to more than one link.
The basic power management mechanism defines two working modes for Wi-Fi devices: active mode and power save (PS) mode \cite{b9}.
In active mode, the device normally transmits and receives frames. In PS mode, the device alternates between the awake state and the doze state to conserve power.


The target wake time (TWT) mechanism was first introduced in IEEE 802.11ah, underwent significant enhancements in IEEE 802.11ax, and was further extended in IEEE 802.11be specifically to support MLO scenarios.
The key part of TWT is the initial negotiation mechanism, in which the TWT session parameters are determined \cite{b7}.
Through the TWT, the STA may exchange frames with the AP during a period of waking up (TWT service periods), and keep dozing out of the TWT service periods. 
Under the MLO architecture, an MLD may negotiate individual TWT agreements with a peer MLD.
Note that the individual TWT agreement is negotiated between the STAs affiliated with the MLDs that are operating on an enabled link and is not negotiated between two MLDs.

\subsection{Low-Latency Operations}
Restricted TWT (R-TWT) operation described in 802.11be enables the STAs in the base station system (BSS) to use enhanced medium access protection and resource reservation mechanisms for delivery of latency sensitive traffic.
R-TWT session period (SP) are announced in the broadcast TWT element and STAs need to establish membership with the R-TWT to use the R-TWT SP.

\begin{table*}[t]
	\renewcommand\arraystretch{1.4}
	\begin{center}
		\caption{Simulation settings with three or four features}
		\label{table_5}
		\setlength{\tabcolsep}{3 mm}{
			\begin{tabular}{|c|c|c|c|c|}
				\hline   \rule{0pt}{8pt} \textbf{Cases} & \textbf{Aggregation} & \textbf{Link} & \textbf{CBW (MHz)} & \textbf{QAM}  \\  
				\hline   \rule{0pt}{8pt} Case 1-1 & 1024 \checkmark & 1 & 320 \checkmark & 4K \checkmark  \\ 
				\hline   \rule{0pt}{8pt} Case 1-2 & 256 & 2 (320 MHz for link 1 + 160 MHz for link 2) \checkmark & 320 \checkmark & 4K \checkmark \\  
				\hline   \rule{0pt}{8pt} Case 1-3 & 256 & 2 (320 MHz for link 1 + 320 MHz for link 2) \checkmark & 320 \checkmark & 4K \checkmark \\   
				\hline   \rule{0pt}{8pt} Case 1-4 & 1024 \checkmark & 2 (160 MHz for link 1 + 160 MHz for link 2) \checkmark & 160 & 4K \checkmark \\ 
				\hline   \rule{0pt}{8pt} Case 1-5 & 1024 \checkmark & 2 (320 MHz for link 1 + 160 MHz for link 2) \checkmark & 320 \checkmark & 1K  \\ 
				\hline   \rule{0pt}{8pt} Case 1-6 & 1024 \checkmark & 2 (320 MHz for link 1 + 320 MHz for link 2) \checkmark & 320 \checkmark & 1K  \\ 
				\hline   \rule{0pt}{8pt} Case 2-1 & 1024 \checkmark & 2 (320 MHz for link 1 + 160 MHz for link 2) \checkmark & 320 \checkmark & 4K \checkmark \\ 
				\hline   \rule{0pt}{8pt} Case 2-2 & 1024 \checkmark & 2 (320 MHz for link 1 + 320 MHz for link 2) \checkmark & 320 \checkmark & 4K \checkmark \\ 
				\hline   \rule{0pt}{8pt} Case 3 & 1024 \checkmark & 3 (320 MHz for link 1 + 160 MHz for link 2 + 160 MHz for link 3 ) \checkmark & 320 \checkmark & 4K \checkmark \\ 
				\hline
		\end{tabular} }
	\end{center}
\end{table*}



\begin{figure}[t]
	\centering
	\includegraphics[width=3.5in]{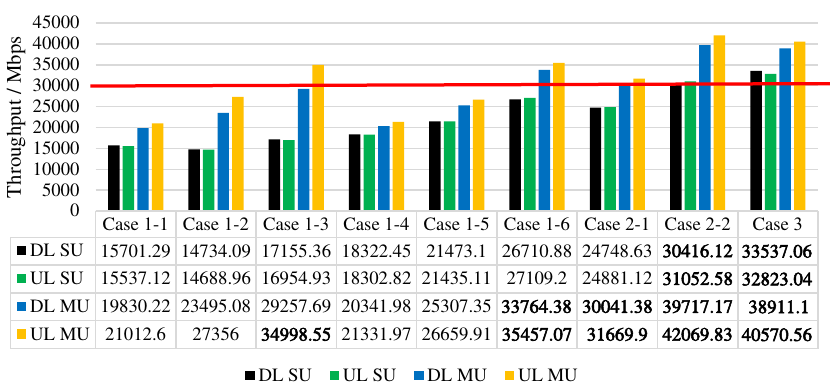}
	\caption{Simulation results of 802.11be throughput (with three or four features).}
	\label{fig_8}
\end{figure}

\section{Performance Evaluation} \label{per}
EHT proposed project authorization request (PAR) requires capable of supporting a maximum throughput of at least 30 Gbps and at least one mode of operation capable of improved the worst case latency and jitter.
In this section, we verify the proposed PAR on latency and throughput through simulations.


\subsection{PAR throughput verification}
In the validation process for achieving 30 Gbps throughput, four features were considered to meet the objectives outlined in the PAR for the project.
These features included multi-link support (with 2 links and STR transmission mode), 4K-QAM modulation, aggregation of 1024 MPDU aggregation per A-MPDU, and a 320 MHz bandwidth.
These features collectively ensure a guaranteed 30 Gbps throughput for both single user (SU) and multi-user (MU) scenarios.
In the simulation scenario, each BSS consists of an AP MLD and $N$ non-AP MLDs placed around it and the throughput is measured at MAC data service access point (SAP).
In terms of simulation setup parameters, the number of non-AP MLD is 8 and the size of area is 20 m $\times$ 20 m. 
The TX power of AP MLD is 20 dBm and the TX power of non-AP MLD is 15 dBm.
The number of links is 2, working at the different frequency bands (i.e., 5 GHz and 6 GHz). We consider both UL and DL full buffer traffic type, which means that there is always data to send, so it is often used to evaluate the maximum throughput of the network.
The complete set of parameters used in the simulation is shown in Table \ref{table_2}.

We first analyzed the simulation results of adding one or two EHT features and compare them with the results of IEEE 802.11ax (i.e., 256 MPDU aggregation for A-MPDU, one link with 160 MHz bandwidth, 1K-QAM). Although the simulation results with one or two features are better than the 802.11ax performance, no single individual feature can reach 30 Gbps, and only combining 320 MHz bandwidth with 2 MLO links for UL MU can achieve this. 
It should be noted that this part of results figures are omitted due to a lack of space.

Second, we consider the combination of three or four EHT features.
The settings of simulation parameters are described in Table \ref{table_5} and the activated features are marked with checkmarks.
It is noteworthy that there are two different ways to combine 320 MHz bandwidth links.
The simulation results are shown in Fig.~\ref{fig_8} and the throughput exceeding 30 Gbps is bolded.
It can be seen that with three combined features, case 1-3 for UL MU, case 1-6 for MU cases achieve 30 Gbps; with four combined features, case 2-1 for MU cases and case 2-2 for both MU and SU cases achieves 30 Gbps.
In addition, it can be concluded that multiple links with 320 MHz can improve throughput.
For example, case 2-2 with four features has 262.77\%, 270.36\%, 373.70\%, 401.76\% throughput gain in DL SU, UL SU, DL SU and UL MU respectively compared with 802.11ax.
In case 3, three links are applied by adding one link with 160 MHz based on case 2-1 and it achieves 30 Gbps for both MU and SU cases.
Therefore, adding links can improve the throughput.

\subsection{PAR low latency verification} \label{lat}
In this section, we verify through simulations that the proposed PAR on latency in 802.11be can be achieved. Through simulation, we can see whether there is an improvement in the delay of the data delivery. In our simulation, the delay is defined as the duration from the time when data is passed to the STA’s MAC layer for transmission, till the reception of its expected immediate response at the STA’s MAC layer.

MLO is considered as one of the most important features in 802.11be and the latency verification will be done using MLO under STR transmission mode to see how much latency gain can be achieved.
The topology in this simulation is based on a residential scenario from the 802.11ax simulation scenarios document \cite{b10}, which is briefly described below.
There are 2 floors (each 3 m high), and 2 $\times$ 10 apartments on each floor.
The apartment size is 10 m $\times$ 10 m $\times$ 3 m and 1/3 of apartments will be using the same channel.
In this simulation, 5 BSSs using the same channel are analyzed with 4 STAs in each BSS.
In every deployment, we will place AP at the center of the BSS and STA distributed randomly in the BSS.
We consider both UL and DL traffic types and the detailed simulation setup parameters are listed in Table \ref{table_2}, where the traffic rate is the packet generation rate for all links at the transmitter.
Next, we set up different simulation cases to compare the latency performance for 802.11ax and 802.11be in DL/UL SU-MIMO and DL/UL OFDMA.

To make this a fair comparison, we set the number of both 802.11ax and 802.11be links to 2, where each link has a bandwidth of 160 MHz.
In the simulation, 5 BSSs are using the same channel.
In 802.11ax, 2 APs works independently on 2 links, each AP has 4 associated STAs. The traffic rate of each AP is $v_{1}$, and the traffic rate of each STA is $v_{2}$.
In 802.11be, 2 APs affiliated with 1 AP MLD work on 2 links, and an AP MLD has 4 associated non-AP MLDs. The traffic rate of each AP MLD is 2$v_{1}$, and the traffic rate of each non-AP MLD is 2$v_{2}$.

\begin{figure}[!t]
	\centering
	\includegraphics[width=3.6in]{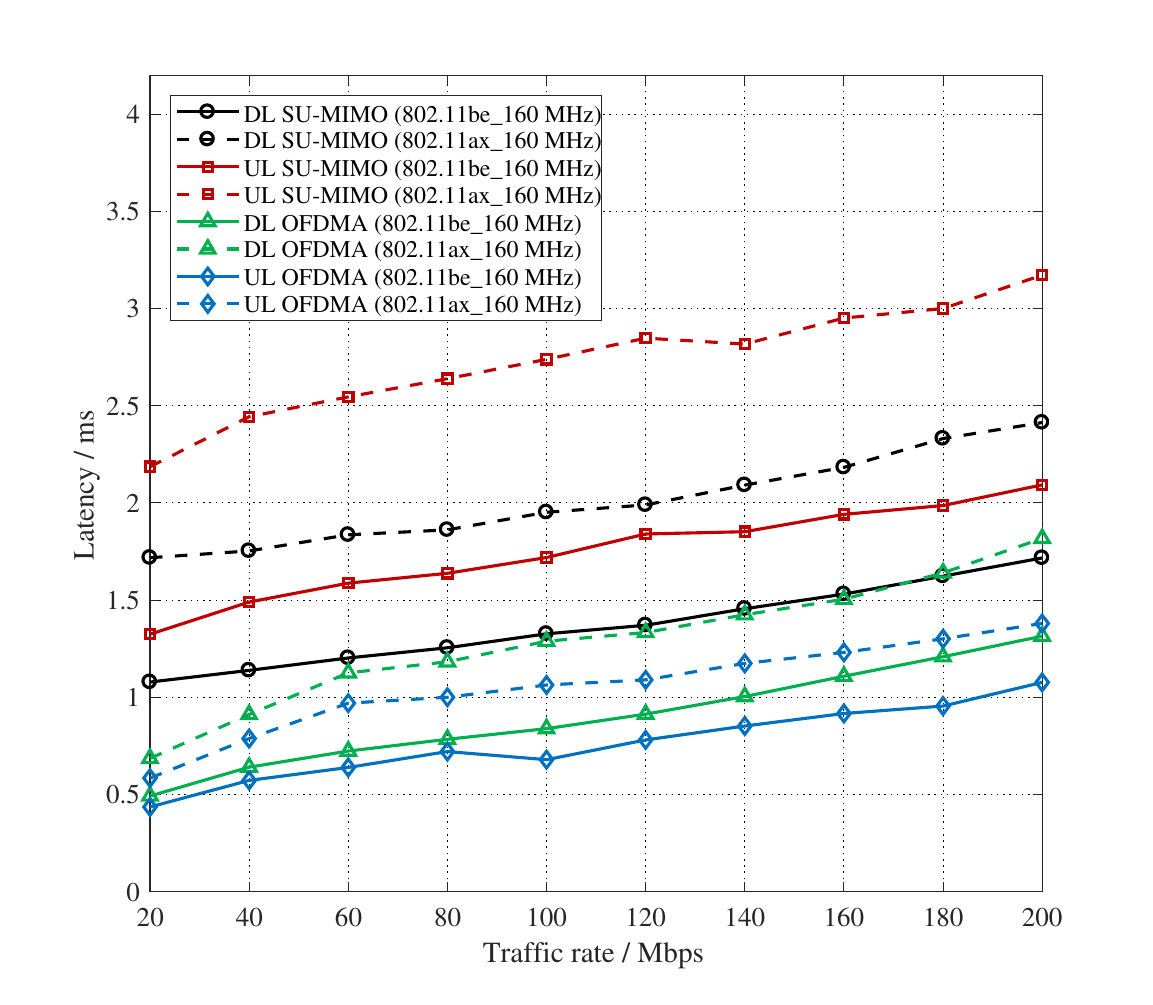}
	\caption{Average delay simulation results of 802.11ax and 802.11be with two 160 MHz links.}
	\label{fig:1234}
\end{figure}

The simulation results are shown in Fig.~\ref{fig:1234},
we can clearly see the latency performance of 802.11be is better in various scenarios.
The average delay gain values of 802.11be compared with 802.11ax are 32\%, 36\%, 30\% and 28\% in DL SU-MIMO, UL SU-MIMO, DL OFDMA and UL OFDMA respectively.
Moreover, regardless of using 802.11ax or 802.11be, the latency performance under OFDMA is superior to SU-MIMO. The reason is that OFDMA allows multiple users to transmit data on different frequency subcarriers at the same time, reducing competition and queuing time among users, thereby decreasing latency.
In contrast, SU-MIMO transmits multiple data streams at the same time and frequency, which may result in queuing up for resources when multiple users try to access the network concurrently.
In terms of scheduling strategies, OFDMA can dynamically adjust the allocation of subcarriers based on the demand of each user and channel conditions.
This flexibility ensures that the system can respond to changes in a short period, providing users with the required resources more swiftly, thereby reducing waiting times and latency.



%
%

\section{Discussion}
In this section, we briefly summarize and discuss the simulation results of low latency and throughput, which is the key objectives of 802.11be.
Our simulations show how much throughput can be achieved by combining features. 
Combining all four features achieves 30 Gbps for both SU and MU cases. Multi-link with two links using 320 MHz channels and 1024 aggregation for the UL MU case can also achieve 30 Gbps throughput. 
Generally, the impact on throughput for EHT features from lowest to highest is: 4K-QAM, 1024 aggregation, 320 MHz, multi-link. 
In addition, our simulations show how much delay improvement can be achieved by applying MLO.
We compare 802.11ax with 2 links and 802.11be with 2 links, and each link has the same bandwidth.
In this case, the delay of 802.11be is reduced around 30\% compared to the delay of 802.11ax.
In addition, compared with SU-MIMO, OFDMA can significantly reduce latency because of its bandwidth resource allocation scheme and flexible scheduling strategy.

\section{Conclusion}

Wi-Fi 7 will lead to extremely high throughput and low latency communications, significantly enhancing real-time applications through digital augmentation.
In this article, we detail the recently updated timeline and key technologies that IEEE 802.11be has adopted on Wi-Fi 7 PHY and MAC, which dominate the improvement of Wi-Fi 7 performance.
We demonstrate that the features of 802.11be can support 30 Gbps throughput and low latency through standard-compliant simulations.
Looking ahead, further research will continue to be studied to lay a foundation for the development of the next generation of Wi-Fi.

%
%

\section*{Acknowledgments}
This research is supported in part by the Shenzhen Natural Science Foundation under Grant JCYJ20200109143016563.


\section{Biography Section}

\begin{IEEEbiographynophoto}{Xiaoqian Liu}
(xq-liu21@tsinghua.org.cn) received the B.S. degree in Communication Engineering from Jilin University, Changchun, China, in 2021, and the M.S. degree in Electronic Engineering from Tsinghua University, Beijing, China, in 2024.
From 2022 to 2023, he was an intern with Huawei Technologies Co., Ltd.
His research interests include Wi-Fi standardization and optical communications.
\end{IEEEbiographynophoto}

\begin{IEEEbiographynophoto}{Yuhan Dong}
(dongyuhan@sz.tsinghua.edu.cn) is with the Shenzhen International Graduate School, Tsinghua University, where he is currently an associate professor and leads the Smart Communication and Computing Group. He received the B.S. and M.S. degrees in electronic engineering from Tsinghua University, Beijing, China, and the Ph.D. degree in electrical engineering from North Carolina State University, Raleigh, NC, USA, in 2002, 2005, and 2009, respectively.
\end{IEEEbiographynophoto}

\begin{IEEEbiographynophoto}{Yiqing Li}
(thisisforwilliam@163.com) received B.E. degree and Master degree in information engineering from Shanghai Jiao Tong University, China. She participated Wi-Fi 7 research in Huawei Technologies Co., Ltd. from 2019 to 2022. She is now in Research Institute of China Telecom Corporation Limited. Her main interests focus on the MAC layer technologies such as channel access methods, multi-link operation and so on.
\end{IEEEbiographynophoto}

\begin{IEEEbiographynophoto}{Yousi Lin}
(linyousi@huawei.com) received the B.S. degree in communication engineering from Shandong University, Shandong, China, in 2016, and the Ph.D. degree from the Department of Electrical and Computer Engineering, Virginia Tech, USA, in 2021. Her current research interests include resource allocation and scheduling, network optimization, joint design of communication and control system in Wi-Fi and cellular networks.
\end{IEEEbiographynophoto}

\begin{IEEEbiographynophoto}{Ming Gan}
	(ming.gan@huawei.com) received the Ph.D degree from University of Science and Technology of China, Hefei, China, in 2014. From September 2012 to September 2013, he was a visiting Ph.D student in Northwestern University, Evanston, IL, USA. He is currently a distinguished engineer in Huawei Technologies Co., Ltd. His research interests includes network coding, compressed sensing, optimization theory, internet of things and WiFi. He is a Voting Member of the IEEE 802.11 Standard Association Working Group. He is an active contributor to 802.11bn (Wi-Fi 8), 802.11be (Wi-Fi 7), 802.11ax (Wi-Fi 6), 802.11ba, full duplex and AI/ML. He was Vice Chair of 802.11 AIML TIG  and is currently serving as Vice Chair of 802.11 AIML SC. 
\end{IEEEbiographynophoto}

\vfill


\begin{thebibliography}{1}
\bibliographystyle{IEEEtran}

\bibitem{b1}
A. Garcia-Rodriguez, D. López-Pérez, L. Galati-Giordano and G. Geraci, ``IEEE 802.11be: Wi-Fi 7 strikes back,'' \textit{IEEE Commun. Mag.}, vol. 59, no. 4, pp. 102-108, Apr. 2021.

\bibitem{b2}
C. Deng et al., ``IEEE 802.11be Wi-Fi 7: New challenges and opportunities,'' \textit{IEEE Commun. Surv. Tutor.}, vol. 22, no. 4, pp. 2136-2166, Jul. 2020.

\bibitem{b3}
T. Adame, M. Carrascosa-Zamacois and B. Bellalta, ``Time-Sensitive Networking in IEEE 802.11be: On the way to low-Latency WiFi 7,'' \textit{Sensor}, vol. 21, no. 5, pp. 4954, Jul. 2021.

\bibitem{b4}
D. Lopez-Perez, A. Garcia-Rodriguez, L. Galati-Giordano, M. Kasslin and K. Doppler, ``IEEE 802.11be Extremely High Throughput: The next generation of Wi-Fi technology beyond 802.11ax,'' \textit{IEEE Commun. Mag.}, vol. 57, no. 9, pp. 113-119, Sep. 2019.

\bibitem{PAR}
L. Cariou, ``802.11 EHT Proposed PAR,'' IEEE 802.11-18/1231r6, Mar. 2019.

\bibitem{laten}
Y. Lin et al., ``CR for PAR low latency verification,'' IEEE 802.11-22/1348r1, Sep. 2022.


\bibitem{b5}
E. Avdotin, D. Bankov, E. Khorov and A. Lyakhov, ``Resource allocation strategies for real-time applications in Wi-Fi 7,'' in \textit{Proc. 2020 IEEE Int. Black Sea Conf. Commun. Netw. (BlackSeaCom)}, May 2020, pp. 1-6.

\bibitem{b6}
``P802.11ax--IEEE Draft Standard for Information technology--Telecommunications and information exchange between systems Local and metropolitan area networks--Specific requirements - Part 11: Wireless LAN Medium Access Control (MAC) and Physical Layer (PHY) Specifications Amendment: Enhancements for Extremely High Throughput (EHT),'' \textit{IEEE P802.11be/D3.0}, Jan. 2023.

\bibitem{bmpdu}
K. Mansour, I. Jabri and T. Ezzedine, ``Revisiting the IEEE 802.11n A-MPDU Retransmission Scheme,'' \textit{IEEE Commun. Lett.}, vol. 23, no. 6, pp. 1097-1100, Jun. 2019.

\bibitem{MLOtcom}
J. Zhang, Q. Tian, Y. Gao, X. Sun and W. Zhan, ``WiFi 7 With Different Multi-Link Channel Access Schemes: Modeling, Fairness and Optimization,'' \textit{IEEE Trans. Commun.}, vol. 72, no. 10, pp. 6225-6236, Oct. 2024.

\bibitem{b8}
E. Khorov, I. Levitsky and I. F. Akyildiz, ``Current status and directions of IEEE 802.11 be, the future Wi-Fi 7,'' \textit{IEEE Access}, vol. 8, no. 1, pp. 88664-88688, May 2020.

\bibitem{b7}
Á. López-Raventós and B. Bellalta, ``Multi-link operation in IEEE 802.11be WLANs,'' \textit{IEEE Wirel. Commun.}, vol. 29, no. 4, pp. 94-100, Aug. 2022.

\bibitem{b9}
K. Huang et al., ``Mutli-link channel access schemes for IEEE 802.11be Extremely High Throughput,'' \textit{IEEE Commun. Std. Mag.}, vol. 6, no. 3, pp. 46-51, Sep. 2022.

\bibitem{b10}
S. Merlin et al., ``TGax Simulation Scenarios,'' IEEE 802.11-14/0980r16, Nov. 2015.

\end{thebibliography}
\end{document}